# Chemical reactivity and magnetism of graphene

**E.F.Sheka\* and L.A.Chernozatonskii\*\***

\* Peoples' Friendship University of the Russian Federation, 117923 Moscow, Russia

\*\* Institute of Biochemical Physics RAS, 19991 Moscow, Russia

**Abstract**. The basic problem of weak interaction between odd electrons in graphene is considered within the framework of broken spin-symmetry approach. The latter exhibits the peculiarities of the odd electron behavior via both enhanced chemical reactivity and magnetism.

## 1. Introduction and Background

Graphene alongside with fullerenes and carbon nanotubes (CNTs) belongs to a particular class of odd-electron carboneous substances that differ from a basic classical benzene system by weakening the interaction between odd electrons. The weakening is a consequence of enlarging length of C-C bonds compared to benzene which causes a partial exclusion of odd electrons from the covalent bonding [1, 2] so that a part of covalently bound $\pi$ electrons of benzene become effectively unpaired. These *effectively unpaired electrons* (EUPEs) provide a partial radicalization of the species as well as a considerable enhancement of their chemical reactivity and magnetism. Once considered for fullerenes [1, 3-5] and single-walled CNTs (SWCNTs) [2, 6], the current paper addresses the problem in graphene.

Weakly interacting electrons with nearly degenerate electron states are considered in the study within the framework of unrestricted broken spin-symmetry (UBS) open-shell approximation. The approach is based on single-determinant wave functions and is well elaborated for both $\Psi$- and $\rho$-based QCh methodologies, basing on unrestricted Hartree-Fock scheme [7] and spin-dependent local-spin-density-approximated DFT [8]. In what follows we shall refer to them as UBS HF and UBS DFT, respectively. UBS HF and UBS DFT find favor over close-shell ones (the latter forms the basis for the "aromaticity concept" that has been widely used when considering both CNTs and graphene until now [9]) because the "correlation effects" by Löwdin [10] are automatically introduced at least through the different-orbital-different-spin understanding. However, if for UBS HF this is the only action towards elaboration of the correlation effects, UBS DFT techniques have additional possibilities concerning functionals in use: from the Fock exchange only functional that causes a severe underestimation of the correlation effects to local density approximation (LDA) that provides a huge overestimation, so mixing the Fock and LDA exchange can provide any beforehand decided result [11]. Another problem connected with DFT concerns the total spin [12]. For long ago has been known that DFT cannot be directly applied to calculation of the spin and space multiplet structure (a comprehensive analysis of problems existing in the present formulation of DFT is performed in [13]). A number of special procedures, that all are beyond the pure DFT scope [14], are suggested to overcome the difficulty. The procedures differ by computation schemes as well as by obtained results so that UBS DFT is method-dependent [11-13]. This greatly complicates a comparison of data obtained by different computational versions and places severe obstacles in computational experiments aimed at obtaining quantitative-structure-activity-relationships (QSARs) attaching them to a particular computational scheme, therewith without a complete sure in the results reliability. From this viewpoint, UBS HF has a certain privilege retaining a transparent single-determinant description, restricted version of which led the foundation of the modern physical and chemistry language, which allows for expressing the necessity of involving electron correlation through conventional terms. On the other hand, the effect of UBS incompleteness of electron correlation is the largest one. This raises the question about practical use of the effect if properly designed.

The first suggestion for using the correlation incompleteness of the UBS solution for practical goals was made thirty years ago [14]. The authors showed that if all electrons spins either are not coupled or have not one direction an extra spin density arises due to Löwdin's symmetry dilemma given rise by unrestricted single-determinant solutions [15]. This is due to spin-contamination character of the latter which causes asymmetry of electron density of UHF solution and asymmetric LDA Hamiltonian with different exchange-correlation potentials for spin-up and spin-down orbitals. The authors suggested attributing this density to EUPEs that are withdrawn from the covalent bonding. Obviously, this was the first indication to enhanced chemical reactivity of the odd-electron systems that was not proved at that time due to lack of the empirical knowledge. The next step was made by Noodleman [16] who suggested using unrestricted spin-contaminated solutions to produce spin-pure ones through the determination of exchange integral $J$ that is directly related to the energy difference of Heisenberg Hamiltonian eigenstates and is of big importance for considering magnetic properties. From this time the term *broken spin-symmetry* solution started. As shown later [1-6, 17], the two main conceptions related to EUPEs and energetic parameter $J$ laid the foundation of the contact between the approximated UBS solutions and physical and chemical reality suggesting a quantified approach to chemical and magnetic properties of odd-electron systems. This approach will be used in the current study applying to graphene.

## 2. Basic relations
### 2.1. Chemical reactivity

Since EUPEs are produced by the spin-contaminated unrestricted solutions they are directly connected with the spin contamination

$$C = \langle \hat{S}^2 \rangle - S(S+1), \qquad (1)$$

where $\langle \hat{S}^2 \rangle$ is the expectation value of the total spin angular momentum that follows from UBS solution. Actually, as shown in [18], the total EUPEs number $N_D$ is expressed as

$$N_D = 2\left( \langle \hat{S}^2 \rangle - \frac{(N^\alpha - N^\beta)^2}{2} \right) \qquad (2)$$

where $N^\alpha$ and $N^\beta$ are the number of electrons with spin α and β, respectively. On the other side, the spin contamination produces an extra spin density (that is particularly evident for singlet state) so that $N_D$ is expressed as a trace of the density [14]

$$N_D = trD(r|r'). \qquad (3)$$

Therefore, to quantify $N_D$ one has to know either $\langle \hat{S}^2 \rangle$ or $trD(r|r')$.

For a single Slater-determinant UBS HF function, the evaluation of both quantities is straightforward since the corresponding coordinate wave functions are subordinated to the definite permutation symmetry so that each value of spin $S$ corresponds to a definite expectation value of energy [12, 13]. Thus, $\langle \hat{S}^2 \rangle$ is expressed as [19]

$$\langle S^2 \rangle = \frac{(N^\alpha - N^\beta)^2}{4} + \frac{N^\alpha + N^\beta}{2} - \sum_{i,j=1}^{NORBS} P_{ij}^\alpha * P_{ij}^\beta. \qquad (4)$$

Here $P_{ij}^{\alpha,\beta}$ are matrix elements of electron density for α and β spins, respectively. Similarly, Ex.(3) has the form [4]

$$N_D = \sum_{i,j=1}^{NORBS} D_{ij}, \qquad (5)$$

where $D_{ij}$ are matrix elements of the spin density expressed as $D = (P^\alpha - P^\beta)^2$. The latter expression is related to the NDDO approximation. The summation in (5) is performed over all atomic orbitals. The atomic origin of the UBS HF function produces another important relation concerning the partitioning of the $N_D$ value over the system atoms

$$N_{DA} = \sum_{i \in A} \sum_{B=1}^{NAT} \sum_{j \in B} D_{ij}, \qquad (6)$$

where $N_{DA}$ is attributed to the EUPEs number on atom $A$. The summation in (6) is performed over all atoms.

As shown [1-6], $N_{DA}$ values are in perfect consistence with the free valence of the atom thus quantifying *atomic chemical susceptibility* (ACS) and highlighting targets that are the most favorable for addition reactions of any type. Successful application of this concept to odd-electron nanocarbons is particularly evident for fullerenes [1, 3-5] and SWCNTs [2, 6]. A correct determination of both $N_D$ and $N_{DA}$ values is well provided by the AM1 (PM3) version of semiempirical UBS HF solution [4] by using CLUSTER-Z1 software [20] used in the current study.

Oppositely to UBS HF, UBS DFT faces a conceptual difficulty in the determination of both $\langle \hat{S}^2 \rangle$ and $trD(r|r')$. This is due to the invariance of the electron density ρ with respect to the permutation symmetry [13] so that DFT does not distinguish states with different spins. All attempts to include the total spin into consideration are connected with either Ψ-based contributions to the DFT body or introducing the spin through exchange and correlation parts of functionals [12, 13]. If spin-dependent exchange potentials can be presented analytically, there is no relation that connects the correlation potential with spin so that its spin-dependence is completely arbitrary [12]. That is why unique DFT relations similar to (4)-(6) are absent so that every casual calculation of either $\langle \hat{S}^2 \rangle$ or $trD(r|r')$ posses of a partial interest and is related to a particular calculation scheme used in the study [21, 22].

### 2.2. Odd electrons magnetism

Molecular magnetism of odd electron systems can be considered in terms of the Heisenberg Hamiltonian [23]

$$\hat{H}_{ex} = JS(S+1), \qquad (7)$$

where $S$ is the total spin while $J$ is the exchange integral or, as mainly referred to now, the magnetic coupling constant [11]. The eigenfunctions of the Hamiltonian are simply spin eigenfunctions, and $J$ is directly related to the energy difference corresponding to these eigenstates. Therefore, determination of the magnetic coupling constant becomes a central point of the magnetism study.

Attempts to apply the Heisenberg description of magnetic interaction to the electronic structure of a molecular electron system have been undertaken by many authors (see reviews [11, 23] and references therein). The successful description of such a delicate physical property lies in the appropriate mapping between the Heisenberg spin eigenstates and the suitable computationally determined electronic states. It is customary to derive the relationship between $J$ and the energy difference of pure spin states.

As for UBS HF approach where electronic states are definitely spin-mapped, the problem consists in the determination of pure spin states and the relevant $J$ value. The problem was perfectly decided by Noodleman [16, 24] within the broken spin-symmetry approach so that in the case of even number of "magnetic" (odd) electrons $J$ is determined as

$$J = \frac{E_{S=0}^{UBSHF} - E_{S_{max}}^{PS}}{S_{max}^2}. \qquad (8)$$

Here $E_{S=0}^{UBSHF}$ and $E_{S_{max}}^{PS}$ are the energy of the UBS HF singlet state and the pure-spin state with maximal spin $S_{max}$ that is the exact single-determinant solution at any calculation scheme, respectively. Consequently, the energy of the pure spin singlet state are determined by the equation [16]

$$E_{S=0}^{PS} = E_{S=0}^{UBSHF} + S_{max} J, \qquad (9)$$

while the energy of the subsequent pure spin states of higher spin multiplicity are determined as

$$E_S^{PS} = E_{S=0}^{PS} - S(S+1)J. \tag{10}$$

As said above, both magnetic coupling constant $J$ and pure spin states cannot be straightforwardly obtained within the DFT scope. Usually, particular procedures are used to reach the goal. Without pretending to give an exhaustive list of publications concerning the problem some representative examples may be found in Refs.[25-29]. Some of these attempts are rather successful in view of comparison with experimental data that is the case of the long-year studying of magnetic behavior of biomoleculat complexes with transition metals [29].

Magnetism is the phenomenon of weak interaction so that the object magnetization proceeds when the $J$ absolute value is small. The smallness of the $J$ value is of a particular importance for the magnetism appearance in systems with the singlet ground state due to the second-order character of the magnetic phenomena in this case [30]. At the same time, the $J$ value obviously, correlates with the EUPEs number and the UBS spin density $D(r|r')$ that both increase when $J$ decreases. However, there is no exact relation between $J$, on one hand, and either $N_D$ or $D(r|r')$, on the other. That is why empirically known upper limit of the absolute $J$ value at the level of $10^{-2}$-$10^{-3}$ kcal/mol [31] cannot be straightforwardly translated into the corresponding values for $N_D$ or $D(r|r')$. So that $J$ retains the only quantity that may quantify the magnetic behavior from the theoretical viewpoint.

## 3. Chemical reactivity of nanographenes

Low and homogeneous chemical reactivity of atoms through over a graphene sheet is usually expected by the predominant majority of scientists dealing with the graphene chemistry. As occurred, it is not the case since the length of equilibrium C-C bonds of graphene exceeds 1.395Å that is the upper limit when the covalent coupling between odd electrons occurs [1, 2]. The calculated results for graphene sheets of different size (nanographens NGrs below) are listed in Table 1. We used rectangular NGrs nominated as ($n_a$. $n_z$) structures following [32]. Here $n_a$ and $n_z$ match the number of benzenoid units on the armchair and zigzag edges of the sheets, respectively (Fig.1). The ACS profile for NGr (**15, 12**) with hydrogen terminated edges presented in Fig.2a demonstrates a rather significant variation of the quantity over atoms. As in the case of fullerenes [1, 4] and SWCNTs [2, 6], $N_{DA}$ values are sensitive to the C-C bond lengths variation. Shown by computations, the equilibrium sheet structures are characterized by varying C-C bond lengths while the structure optimization in all cases starts with constant bond lengths through over the sheet with the C-C bond length of 1.42A. As seen from the figure, the highest ACS are characteristic of carbon atoms at the zigzag edges, that is caused by the largest distance between the contour C atoms while those of the armchair edges are similar to ACS values over sheets and are comparable with the ones of SWCNTs sidewall.

When hydrogen terminators are removed, the ACS profile over the sheet remains unchanged while $N_{DA}$ values on both zigzag and armchair edges grow significantly (Fig.2b) still conserving bigger values in favor of zigzag edges.

Obtained results make allowance for the following conclusions concerning chemical reactivity of NGR.

- Any chemical addend will be first of all attached to NGr zigzag edges, both hydrogen terminated and empty;
- Slightly different by activity armchair edges of non-terminated NGrs will compete with zigzag edges;
- Enhanced chemical reactivity of edges atoms allows NGr for behaving as a molecular entity that will willingly attach another spatially enlarged molecular object such as either CNT or substrate surface following with a particular NGr orientation to the objects (see CNT-NGr composites in [33]);
- Chemical reactivity of inner atoms does not depend on the edge termination and is comparable with that of SWCNT sidewall [2, 6] and fullerenes [1, 4] thus providing a range of addition reactions at NGr surface. Recent direct observation of spontaneous chemical adsorption of individual hydrogen and carbon atoms on inner atoms of graphene sheets [34] is well consistent with the prediction followed from the current UBS HF calculations.

Let us look at the results of the UBS DFT studies of NGrs. First notification about peculiar edge states of graphene ribbons appeared as early as 1987 [35] but further extended study started about ten years later [36, 37]. Since that time three main directions of the peculiarity investigation have been shaped up that were focused on 1) edge states within band structure of graphene; 2) chemical reactivity of graphene's ribbon zigzag edges; and 3) magnetism caused by zigzag edges of graphene ribbons. The first topic lays mainly within the framework of the solid state theory concerning the formation of localized states caused by the breakage of translational symmetry in a certain direction that occurs when a graphene sheet is cut into graphene ribbons. This fundamental property is well disclosed computationally independently on the technique used [36, 37] and have approved experimentally [38, 39]. Two other topics are intimately connected with UBS DFT [41-45] that revealed open-shell character of the graphene singlet ground state.

The first UBS DFT examination of chemical reaction between hydrogen-terminated NGrs and common radicals [41] has revealed *unpaired π electrons* distributed over zigzag edges of 0.14$e$ in average on each atom ($N_{DA}$ in terms of the current paper ). This allowed the authors for making conclusion about special chemical reactivity of the atoms as well of their partial radical character (compare the statement with one of the first made for fullerenes [3]). The next authors' conclusion concerns nonedge ribbon carbon atoms, armchair atoms as well as CNT (presumably sidewall) atoms that show little or no radical character.

The cited UBS DFT results correlate with those of UBS HF of the current study in two points. Both approaches disclose open-shell character of the ground singlet state of graphene and establish the EUPEs availability. However, the EUPEs numbers differ by an order of magnitude that restricted the UBS DFT discussion of the chemical reactivity of graphene by zigzag edge atoms only. This is the very point where the difference between UBS HF and UBS DFT approaches is clearly exhibited. The fixation of the open-shell character of the NGR ground singlet state by both UBS techniques is obvious due to single-determinant character of wave functions in the two cases. The open-shell character has revealed due to considerable weakening of the odd electron interaction in graphene caused by rather large C-C bond lengths. As for the magnitude of the unpaired odd ($\pi$) electrons numbers $N_{DA}$, it is difficult to discuss the DFT value since no indication of the way of its determination is presented. Its decreasing by one order of magnitude comparing with UBS HF data might indicate the pressed-by-functional character of the UBS DFT calculations. At any rate, the results clearly exhibit much lower sensitivity of the UBS DFT approach to chemical reactivity of atoms that can be imagined as lifting zero reading level up to 0.2-0.3$e$ in Fig.2a and up to 1.1$e$ in Fig.2b, thereafter the fixation of values below the level becomes impossible. Close-to-zero chemical reactivity of NGrs inner atoms predicted by UBS DFT calculations strongly contradicts to active chemical adsorption of individual hydrogen and carbon atoms on NGr surface recently disclosed experimentally [34].

Difference in the two approaches sensitivity to chemical reactivity can be traced by comparison of bond dissociation energies (BDEs) calculated by using UBS DFT [41] and UBS HF techniques. Table 2 lists the obtained data alongside with experimental findings. As seen from the table, the UBS HF BDEs are remarkably closer to experimental data just demonstrating a stronger ability of the approach in tackling delicate chemical reactivity features.

## 4. Magnetism of zigzag edged nanographenes

The phenomenon, predicted and studied computationally for NGrs, is one of the hottest issue of the NGR science (enough to mention tens of papers presented on recently hold International conference on Nanotubes Science NT08 [9]). At the heart of the statement of the NGR magnetism are localized states whose flat bands are located in the vicinity of the Fermi level and whose peculiarities were attributed to zigzag edges [36-38, 43-46]. In numerous theoretical studies (see Ref. [46] and references therein), this fact was connected with magnetism through focusing on the spin density on edge atoms. Computations were carried out in presumably $\Psi$-contaminated UBS DFT approximations following such a logical scheme: taking into account spins of edge atoms at the level of wave function; considering so-called antiferromagnetic (AFM) and ferromagnetic (FM) spin configurations with spin alignment up on one edge and down (up) on the other edge, respectively, or nonmagnetic configuration when up-down spin pairs are located at each edge, and performing calculations for these spin configurations. The obtained results have shown that 1) the AFM configuration corresponds to the ground state and is followed in stability by FM and then nonmagnetic states; 2) the calculated spin density on edge atoms corresponds to the input spin configurations in all cases. It should be added that numerical results obtained in different studies differ from each other when different functionals were used under calculations. Not taking the singlet multiplicity of the ground state into account, the AFM configuration followed by spin density at zigzag edge atoms is accepted as a decisive point in heralding magnetism of graphene ribbons after which the phenomenon is considered to be confirmed that gives rise to a big optimism towards the expectation of a number of exciting applications of the material, in spintronics for example [47]. This is just a wrong conclusion whose danger was forestalled by Illas et.al. [25]: "…will lead to the absurd consequence that two equivalent BS approach, such as UHF and DFT, will have to use two different ways to compute the magnetic coupling constant. By bringing this argument to the limit one will get the absurd conclusion that the antiferromagnetic state of a periodic system leads to the pure singlet state". However, the UBS DFT singlet state is as spin-contaminated as the UBS HF one and the availability of the spin density is just a strong confirmation of the contamination.

Computed by using UBS HF approach, the spin density distribution over NGr (**15,12**) atoms with hydrogen-terminated and empty edges is demonstrated in Fig.3. Oppositely to the UBS DFT results, spin density is available at all atoms of the graphene sheet. Summing it over the atoms gives zero in both cases, which corresponds to the singlet ground state. As seen from the figure, in both cases spin density at zigzag edge atoms is the highest, even absolutely dominating in the second case. To see only these atoms means lifting the zero reading level up (down) to

~±0.4 in the first case and to ±1.3 in the second one, that, by other words, means lowering sensitivity in recording the density values. This is the same situation caused by the pressed-by-functional character of the UBS DFT solution that was discussed for the ACS profiles in the previous section. It should be noted that spin density on zigzag edge is distributed quite peculiarly not following the above mentioned AFM configuration predicted for the ground state by UBS DFT. However, presented in Fig.3 corresponds to the lowest energy and seems quite natural since the number of odd electrons participating in the formation of the spin density distribution is not restricted to the number of zigzag edge atoms but equals to the total number of atoms of the NGR sheet. Remaining that spin density value is sensitive to the C-C bond length, it becomes clear why varying the latter produces variation in the density distribution as well.

Coming back to magnetism of graphene ribbons, we have to stand from the fact that real ground state of the object is pure-spin singlet. This means that real spin density at each atom is zero. We can nevertheless discuss the possibility of the magnetic behavior of the object however not from the spin density viewpoint but addressing the energy difference between states of different spin multiplicity as was discussed in Section 2.2.

Computed accordingly to Eq. (8)-(10) $E_{S=0}^{UBSHF}$, $J$, and $E_{S=0}^{PS}$ values related to the studied NGrs are listed in Table 3. As seen from the table, the ground state of all species is singlet so that a question arises if the magnetism for singlet ground state object is possible. As discussed in [30], the phenomenon may occur as a consequence of mixing the state with those of high multiplicity following, say, to van Fleck mixing promoted by applied magnetic field [48]. Since the effect appears in the first-order perturbation theory, it depends on $J$ that determines the energy differences in denominators. Consequently, $J$ should be small to provide noticeable magnetization. Obviously, singlet-triplet mixing is the most influent. As follows from Table 3, the energy gap to the nearest triplet state for the studied NGrs constitutes 1-4 kcal/mol. The value is large to provide a noticeable magnetization of a molecular magnet [31]. However, the value gradually decreases when the number of odd electrons increases. The behavior is similar to that obtained for fullerene oligomers [5] that led to the suggestion of a scaly mechanism of nanostructured solid state magnetism of the polymerized fullerene $C_{60}$.

In view of this idea, let us estimate how large should be NGr to provide a noticeable magnetization. As known [31], molecular magnetism can be fixed at $J$ value of $10^{-2}$-$10^{-3}$ kcal/mol. Basing on the data presented in Table 2 and supposing the quantity to be inversely proportional to the number of odd electrons $N$, we get $10^3 - 10^4$ for the latter. In NGrs $N$ coincides with the number of carbon atoms that is determined for rectangular NGrs as [32]

$$N = 2(n_\alpha n_z + n_\alpha + n_z), \tag{11}$$

where $n_a$ and $n_z$ are the numbers of benzenoid units on the armchair and zigzag ends of the sheets, respectively. To fit the needed $N$ value, the indices $n_\alpha$ and $n_z$ should be of tens to hundreds that leads to linear sizes of the NGrs of a few nm. The estimation is rather approximate but it nevertheless correlates well with experimental observations of the magnetization of activated carbon fibers consisting of nanographite domains of 3-5 nm in size [49, 50].

## 5. Conclusion

The electronic behaviour of graphene is governed by the interaction of odd electrons that is much weaker than, say, in ethylene and benzene due to which a lot of nearly degenerate states appear in the energy spectrum. As a consequence, a theoretic description of electronic properties, particularly in singlet state, must include configurational interaction which greatly complicates the computational analysis. As known, broken spin-symmetry approach is one of effective approximations to tackle the problem. Modern implementations of the approach in the form of either unrestricted Hartree-Fock scheme or unrestricted DFT were discussed with particular attention to the applicability of spin-contaminated solutions of both techniques for the description of electronic properties of graphene. While UBS DFT is limited to revealing open-shell character of the singlet state of the object only, UBS HF demonstrates a unique sensitivity in revealing both enhanced chemical activity and magnetism of graphene sheets and/or ribbons. The former is presented in terms of a quantified atomic chemical susceptibility that is homogeneously distributed over all nonedge atoms with the value similar to that of fullerenes and CNTs sidewalls. Peculiarities of chemical reactivity of graphene sheets are connected with edge atoms that show twice and five times increasing of the susceptibility for zigzag edge atoms in the case of either terminated or empty edges, respectively. Armchair edge atoms prevail over inner ones only in the absence of chemical termination of the edges. Magnetic response of graphene sheets is shown to be size-dependent due to lowering magnetic coupling constant and approaching the latter to a limit value when the sheet size increases. The limit value is attained when linear sizes of a sheet are of a few nm that is well consistent with experimental findings.

The explanation suggested by the UBS HF approach seems quite reasonable. A common view on both chemical reactivity and magnetism of graphene, physically clear and transparent, witnesses the approach internal self-consistency and exhibits its high ability to quantitatively describe practically important consequences of weak interaction between odd electrons.

**ACKNOWLEDGMENTS**. The work was supported by the Russian Foundation for Basic Research (grant 08-02-01096).

**Table 1**. Atomic chemical susceptibility of H-terminated nanographenes

| Nanographenes $(n_a, n_z)$[1] | $N_{DA}$ | | |
|---|---|---|---|
| | Armchair edge | Central part | Zigzag edge |
| (15, 12) | 0.28-0.14 | 0.25-0.06 | 0.52-0.28 |
| (15, 12)[2] | 1.18-0.75 | 0.25-0.08 | 1.56-0.93 |
| (7, 7) | 0.27-0.18 | 0.24-0.12 | 0.41-0.28 |
| (5, 6) | 0.27-0.16 | 0.23-0.08 | 0.51-0.21 |

[1] Following [32], $n_a$ and $n_z$ match the numbers of benzenoid units on the armchair and zigzag ends of the sheets, respectively.
[2] After removing hydrogen terminators

**Table 2**. Comparison of bond dissociation energy (BDE, in eV) of zigzag edge-X bonds with experimental BDE of $C_2H_5$-X. Zero-point-vibration energy corrections are not included

| Radical X | H | OH | $CH_3$ | F | Cl | Br | I |
|---|---|---|---|---|---|---|---|
| BDE (edge X) UBS DFT[1] | 2.86 | 2.76 | 2.22 | 3.71 | 2.18 | 1.65 | 1.18 |
| BDE (edge X)[2] UBS HF | 4.27 | 3.86 | 3.42 | 4.50 | 2.25 | - | - |
| BDE ($C_2H_5$-X)[3] | 4.358 | 4.055 | 3.838 | 4.904 | 3.651 | 3.036 | 2.420 |

[1] Data from Ref.[41]
[2] The current study calculations were performed for NGR (**5, 6**).
[3] Experimental values from Ref. [42].

**Table 3.** NGrs electronic characteristics in *kcal/mol*

| Nanographenes[1] | The number of "magnetic"(odd) electrons | $E_{S=0}^{UBSHF}$ | $J$ | $E_{S=0}^{PS}$ | Singlet-triplet gap[2] |
|---|---|---|---|---|---|
| (15, 12) | 400 | 1426.14 | -0.42 | 1342.14 | 0.84 |
| (7, 7) | 120 | 508.69 | -1.35 | 427.69 | 2.70 |
| (5, 6) | 78 | 341.01 | -2.01 | 262.72 | 4.02 |

[1] Nomenclature of nanographenes is given in Footnote 2 to Table 1.
[2] For pure-spin states the singlet-triplet gap $E_{S=1}^{PS} - E_{S=0}^{PS} = -2J$ [16].

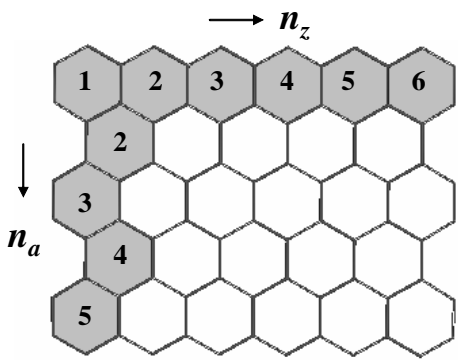

**Figure 1.** Notification of a rectangular NGr ($n_a$. $n_z$) with armchair ($n_a$) and zigzag ($n_z$) edges [32]

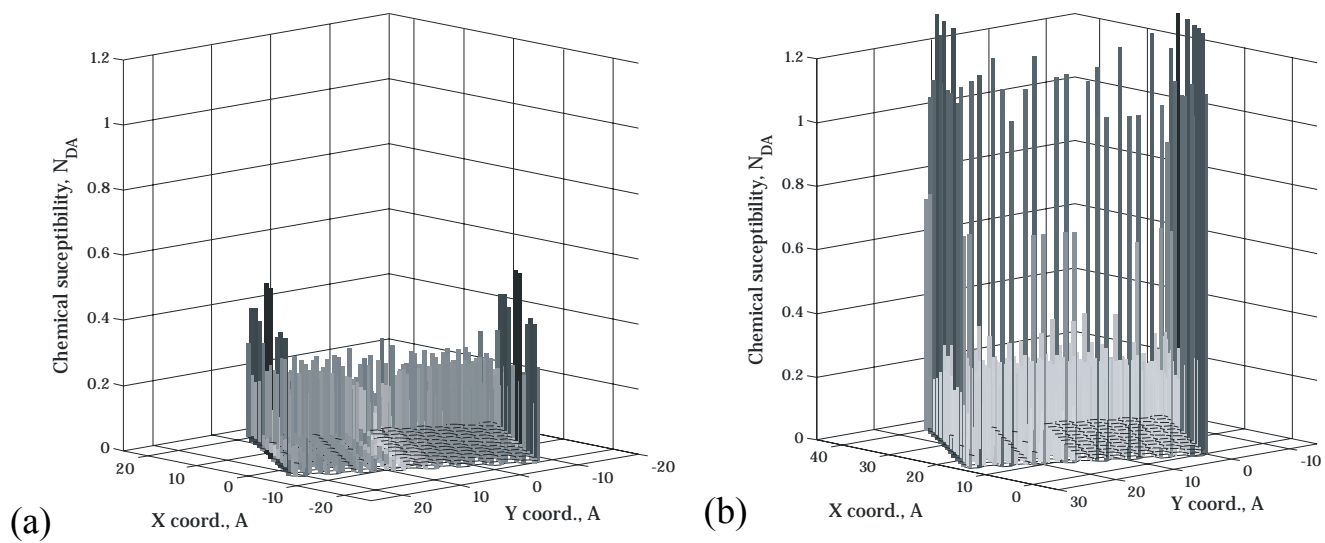

**Figure 2.** Distribution of atomic chemical susceptibility over atom of rectangular NGr (**15,12**) with hydrogen terminated (a) and empty (b) edges. UBS HF solution. Singlet state.

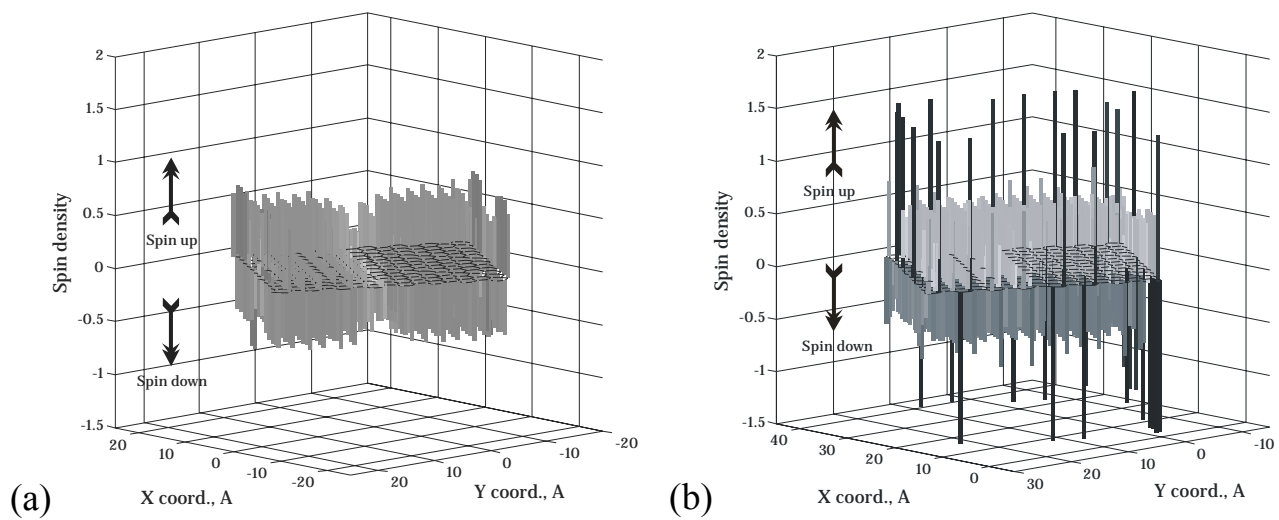

**Figure 3.** Distribution of spin density over atom of rectangular NGr (**15,12**) with hydrogen terminated (a) and empty (b) edges. UBS HF solution. Singlet state.